\documentclass[fleqn,usenatbib]{mnras}

\usepackage{newtxtext,newtxmath}

\usepackage[T1]{fontenc}

\DeclareRobustCommand{\VAN}[3]{#2}
\let\VANthebibliography\thebibliography
\def\thebibliography{\DeclareRobustCommand{\VAN}[3]{##3}\VANthebibliography}

\usepackage{graphicx}	
\usepackage{amsmath}	
\usepackage[dvipsnames]{xcolor}
\usepackage{comment}
\usepackage{tablefootnote}
\usepackage{soul}
\usepackage[normalem]{ulem}
\usepackage{xcolor}
\usepackage{soul}
\usepackage{pifont}

\newcommand{\DI}[0]{D\,{\sc i}}
\newcommand{\HI}[0]{H\,{\sc i}}
\newcommand{\OI}[0]{O\,{\sc i}}
\newcommand{\CII}[0]{C\,{\sc ii}}
\newcommand{\SiII}[0]{Si\,{\sc ii}}
\newcommand{\AlII}[0]{Al\,{\sc ii}}
\newcommand{\FeII}[0]{Fe\,{\sc ii}}
\newcommand{\q}{J\,1332$+$0052}
\newcommand{\ql}{SDSS~J133254.51+005250.6}
\newcommand{\ioffe}{Ioffe Institute, {Polyteknicheskaya 26}, 194021 Saint-Petersburg, Russia}
\newcommand{\eso}{European Southern Observatory, Alonso de C\'ordova 3107, Vitacura, Casilla 19001, Santiago, Chile}
\newcommand{\iap}{Institut d'Astrophysique de Paris, CNRS-SU, UMR\,7095, 98bis bd Arago, 75014 Paris, France}
\newcommand{\fcla}{Franco-Chilean Laboratory for Astronomy, IRL 3386, CNRS and U. de Chile, Casilla 36-D, Santiago, Chile \label{fcla}}
\newcommand{\sut}{Centre for Astrophysics and Supercomputing, Swinburne University of Technology, Hawthorn, VIC 3122, Australia\label{sut}}

\title[A New Precise Determination of the Primordial Abundance of Deuterium.]{A New Precise Determination of the Primordial Abundance of Deuterium:
Measurement in the metal-poor sub-DLA system at $z=3.42$ towards quasar \q\thanks{Based on archival data collected at the European Southern Observatory under programme 0101.A-0061(A)
and W.M. Keck Observatory under programmes U152Hr, U088Hb, and U088Hr.}
}

\author[Kislitsyn P. et al.]{
P.A.~Kislitsyn$^{1}$\thanks{E-mail: pavel.kislitsyn@gmail.com},
S.A.~Balashev$^{1}$,
M.T.~Murphy$^{2}$,
C.~Ledoux$^{3}$,
P.~Noterdaeme$^{4,5}$
and A.V.~Ivanchik$^{1}$
\\
$^{1}$ \ioffe \\
$^{2}$ \sut \\
$^{3}$ \eso \\
$^{4}$ \fcla \\
$^{5}$ \iap \\
}

\date{Accepted 2024 January 18. Received 2024 January 18; in original form 2023 September 30}

\pubyear{2023}

\begin{document}
\label{firstpage}
\pagerange{\pageref{firstpage}--\pageref{lastpage}}
\maketitle

\begin{abstract}

The theory of Big Bang nucleosynthesis, coupled with an estimate of the primordial deuterium abundance (D/H)$_{\rm pr}$, offers insights into the baryon density of the Universe. Independently, the baryon density can be constrained during a different cosmological era through the analysis of cosmic microwave background (CMB) anisotropy. The comparison of these estimates serves as a rigorous test for the self-consistency of the Standard Cosmological Model and stands as a potent tool in the quest for new physics beyond the Standard Model of Particle Physics.
For a meaningful comparison, a clear understanding of the various systematic errors affecting deuterium measurements is crucial. Given the limited number of D/H measurements, each new estimate carries significant weight. This study presents the detection of D\,{\sc i} absorption lines in a metal-poor sub-Damped Lyman-$\alpha$ system ($\rm [O/H]=-1.71\pm 0.02$, $\log N($\HI$)=19.304\pm 0.004$) at $z_{\rm abs}=3.42$ towards the quasar \ql. Through simultaneous fitting of \HI\ and \DI\ Lyman-series lines, as well as low-ionization metal lines, observed at high spectral resolution and high signal-to-noise using VLT/UVES and Keck/HIRES, we derive $\log($\DI/\HI$)=-4.622\pm 0.014$, accounting for statistical and systematic uncertainties of $0.008$~dex and $0.012$~dex, respectively.
Thanks to negligible ionization corrections and minimal deuterium astration at low metallicity, this D/H ratio provides a robust measurement of the primordial deuterium abundance, consistent and competitive with previous works. Incorporating all prior measurements, the best estimate of the primordial deuterium abundance is constrained as: (D/H)$_{\rm pr}=(2.533\pm 0.024)\times 10^{-5}$. This represents a 5\% improvement in precision over previous studies and reveals a moderate tension with the expectation from the Standard Model ($\approx$2.2$\sigma$). This discrepancy underscores the importance of further measurements in the pursuit of new physics.
\end{abstract}




\section{Introduction}



Primordial nucleosynthesis is one of the pillars of the Big Bang cosmology that provides the earliest robust observational constraint on the Standard Cosmological Model. The theory of primordial nucleosynthesis predicts the values of the relative abundances of light nuclei such as H, D, $\rm ^3He$, $\rm ^4He$, and $\rm ^7Li$.  According to the Standard Model, these abundances depend only on the baryon-to-photon ratio, $\eta_{\mathrm{b}} \equiv \frac{n_{\mathrm{b}}}{n_\gamma}$ (where $n_{\mathrm{b}}, n_\gamma$ -- the number densities of baryons and photons, respectively), that is linearly dependent on the baryon density, $\Omega_{\mathrm{b}}$. However, some of these nuclei are sensitive to effects beyond the Standard Model. As an example, the primordial helium abundance could provide better constraints on the effective number of neutrino species \citep{Kurichin2021}. This draws a lot of attention to the determination of the primordial helium abundances and its estimations are steadily improving \citep[e.g.,][]{Izotov_2014, Fernandez, Hsyu_2020, Aver2021, Kurichin2021_2, Valerdi, EMPRESS_2022}. Despite growing observational constraints on helium, deuterium plays a key role in the modern era of precise cosmology since it has the strongest sensitivity on $\Omega_{\mathrm{b}}$ among the other elements and its abundance can be accurately constrained using absorption-line measurements along quasar sightlines.


With the advent of high-resolution spectrographs on the largest optical telescopes, for more than two decades D/H measurements were derived from the analysis of hydrogen and deuterium absorption lines in quasar spectra \citep{Tytler_1996, Noterdaeme_2012, Cooke_2014, Cooke_2018}. The first results on D/H possessed a high dispersion, the reasons of which could be explained by systematic effects in the analysis or could have physical origins. This has motivated the search for criteria that minimize the systematics and provide the most precise D/H measurements \citep{Cooke_2014}. The main criteria are a simple velocity structure for the absorber, low deuterium astration and depletion on the dust \citep{Cooke_2018}, which are all satisfied in case of very low metallicities. Therefore, only a few D/H estimations are considered to date to provide robust results \citep{Cooke_2018}. 
Each new estimate is of utmost importance, for various combinations of parameters (e.g., metallicity, HI column density, simplicity of the velocity structure, redshift), that allow to disentangling the influence of the aforementioned systematic effects on the measurement of D/H. Such measurements are especially important at the high-metallicity end, which can potentially increase the number of suitable targets, especially in the advent of next-generation spectrographs such as CUBES \citep{Cubes2022}.

In this paper, we present the detection and analysis of \DI\ lines in the sub-DLA system at $z_{\rm abs}=3.42$ towards \ql. This provides a new robust determination of the D/H ratio at high redshift.

\section{Observations and data reduction}
\label{sec:obs}

\subsection{VLT/UVES observations}

\ql\ (hereafter J\,1332$+$0052), a $z_{\rm em}=3.51$, $V=18.7$ quasar (a.k.a. Q\,1330$+$0108; \citealt{Veron-Cetty2010}) was observed in service mode between April and July 2018 using the Very Large Telescope (VLT) Unit-2, Kueyen, equipped with the Ultraviolet and Visual Echelle Spectrograph \citep[UVES;][]{UVES}. Twelve exposures of 3000~s each were taken in Dichroic mode over seven different nights. This resulted in 24 individual spectra covering most of the optical range. Information about instrumental setups and observing conditions is given in Table~\ref{tab:uves_exposures}. Entrance $1\arcsec$-wide slits and $2\times 2$ pixel binning were used throughout. The observations were carried out in dark time (no moon) under clear skies and excellent seeing conditions (see table~\ref{tab:uves_exposures}).
For each exposure, we processed science and calibration data using the UVES pipeline v6.1 on the EsoReflex platform v2.11 \citep{Freudling2013}. Intermediate data products were inspected carefully and data-reduction parameters were optimised step by step. The nominal resolving power of the individual spectra is 50,000 in the Blue (48,800 in the Red) but can be slightly larger than that when the seeing conditions are better than $1\arcsec$ FWHM. The actual resolving power of the combined spectrum used in the analysis is determined in Sect. \ref{sect:fit}.
Individual exposures in each blue/red setup were co-added optimally and the overlapping regions of the resulting spectra were stitched together to create the final data product.
The final UVES spectrum has a fairly high S/N ratio per pixel, peaking at $\approx 30$ in the quasar continuum around the most important \DI\ lines.

\begin{table}
\centering
\caption{VLT/UVES observations of \q}
\label{tab:uves_exposures}
\begin{tabular}{ccccc} 
\hline
UT date    & exp. time       & setup$^{\rm a}$ & airmass        & IQ$^{\rm b}$ \\
           & [s]             & [nm]            &                & [$\arcsec$]           \\
\hline
09-04-2018 & $2 \times 3000$ & 437+760         & 1.27           & 0.83                 \\
13-04-2018 & $4 \times 3000$ & 437+760         & 1.12/1.16      & 0.55/0.6             \\
18-05-2018 & $2 \times 3000$ & 390+564         & 2.02           & 1.24                  \\
12-06-2018 & $4 \times 3000$ & 390+564         & 1.23/1.44      & 0.98/1.05             \\
13-06-2018 & $6 \times 3000$ & 437+760         & 1.19/1.36/1.70 & 0.78/1.08/1.22        \\
15-06-2018 & $4 \times 3000$ & 437+760         & 1.11/1.14      & 1.09/1.17             \\
09-07-2018 & $2 \times 3000$ & 390+564         & 1.70           & 1.23                  \\
\hline
\end{tabular}
\flushleft $^{\rm a}$ central wavelengths of the blue-arm (left) and red-arm (right) spectra.\\
$^{\rm b}$ image quality at $\lambda=650$~nm measured by the telescope Shack-Hartmann wavefront sensor.
\end{table}

\subsection{Keck/HIRES observations}

In addition to the UVES spectra, we used archival spectra from the W.M. Keck Observatory using the High-Resolution Echelle Spectrometer \citep[HIRES;][]{Vogt1994}. The journal of observations is given in Table~\ref{tab:keck_exposures}. The nominal resolving power of the individual exposures is 47,700.

To reduce the data (eight separate exposures) and combine the extracted spectra, we used the approach described in \cite{Robert_2019}. To summarise it briefly, the data were reduced and the quasar spectra extracted using the {\sc makee}\footnote{See \url{http://www.astro.caltech.edu/~tb/makee}\,.} package, including corrections for the blaze function using flat-field exposures and wavelength calibration from ThAr lamp exposures. The extracted spectra from all echelle orders, and all exposures, were combined into a nearly continuous 1-dimensional spectrum using the {\sc uves\_popler} software (\citealt{Murphy_2016}; \citealt{Murphy_2019}). All spectra were redispersed onto the same wavelength grid with 2.2\,km\,s$^{-1}$ pixels and combined using an inverse-variance-weighted mean. An initial continuum was automatically fitted to the regions redwards of the Lyman-$\alpha$ emission line, whilst a manually-fitted polynomial was used for the bluer regions affected by the Lyman-$\alpha$ forest.
The combined HIRES spectrum has a S/N ratio per pixel of $\approx 20$ in the quasar continuum around the most important \DI\ lines.

\begin{table}
\centering
\caption{Keck/HIRES observations of \q}
\label{tab:keck_exposures}
\begin{tabular}{ccccc} 
\hline
UT date    & exp. time       & wavelength range & airmass        & decker \\
           & [s]             &    [nm]          &                &         \\
\hline
07-04-2006 & $2 \times 3600$ & 424-869         & 1.08/1.06      & C1             \\
12-04-2015 & $4 \times 3600$ & 392-686         & 1.06 - 1.20    & C1\\ 
14-05-2015 & $3288+2678$     & 424-869         & 1.15/1.07      & C1             \\
\hline
\end{tabular}
\end{table}

\section{Analysis}
\subsection{A method} 
\label{sect:method}

To determine the properties of the sub-DLA system, we used joint multi-component Voigt-profile fitting\footnote{We used the python package spectro (\url{https://github.com/balashev/spectro}).} of hydrogen, deuterium, and low-ionisation metal absorption lines. Within this approach, each considered absorption line is described by a collection of components, that are defined by the redshift, Doppler parameter, and column density of each species. We used the same number of components for each line (except for \FeII\ where only the two main components are detected), with redshifts tied between lines. We also assumed that the Doppler parameters of each species within one component are tied based on the assumption of micro-turbulence, where the velocity distribution of each species is described by a Gaussian function with Doppler parameter

\begin{equation}
 b_{\rm sp} = \sqrt{b_{\rm turb}^2 + \frac{2 k_B T}{M_{\rm s}}},
\end{equation}

where $k_B$ is the Boltzmann constant, $M_{\rm s}$ is the atomic mass of species $s$, $T$ is the kinetic temperature, and $b_{\rm turb}$ is the turbulent broadening parameter. Consequently, the latter two parameters were fitted independently in each velocity component.

To make the column-density determinations trustworthy, we carefully inspected each portion of the spectra to which the fitting procedure was applied. We simultaneously fitted a
ny unrelated absorption which might have an impact on the fit. We also paid special attention to including all available \DI\ and \HI\ lines to avoid unconscious bias leading to include/exclude lines which may seem to be more/less suitable.

For the metal lines and Lyman-series lines, we first reconstructed a local continuum by spline interpolation from regions without evident absorption. While for the metal lines, outside Ly$\alpha$ forest, we found that continuum can be well constrained, for some of Ly-series lines it is not the case, and assumed continuum may impact on the derived \HI\ and \DI\ column densities \citep[see, e.g.,][]{Balashev_2016}. Therefore we fitted the continuum for several selected Lyman-series lines simultaneously with line profiles to exclude a bias related to a manual continuum placement.  
To do this we used Chebyshev polynomials of order five, three and one in the regions 
near Ly$\alpha$, Ly$\beta$, other Lyman-series lines, respectively.
In addition, we fitted the effective resolutions of the final UVES and HIRES spectra as independent parameters.

We constrained the probability density distribution of the fit parameters within the Bayesian framework, using the Monte Carlo Markov Chain technique with Affine invariant sampler \citep{sampler}. We assumed flat priors on $b_{\rm turb}$, $T_{\rm kin}$, $z$, $\log N$, the continuum Chebyshev coefficients, $C_{i}$, and resolutions of the spectra $R_{\text{UVES}}$, $R_{\text{HIRES}}$, with enough width to guarantee independence of the posteriors on it. We also assumed a same D/H value for all components, i.e. we used a single parameter for the whole subDLA. To report the resulting fit parameters and their uncertainties, we used the maximum posterior probability estimates and highest posterior density 68.3\% credible intervals, respectively.

\subsection{Fit description and results}
\label{sect:fit}

\begin{figure*}
	\includegraphics[width=2\columnwidth]{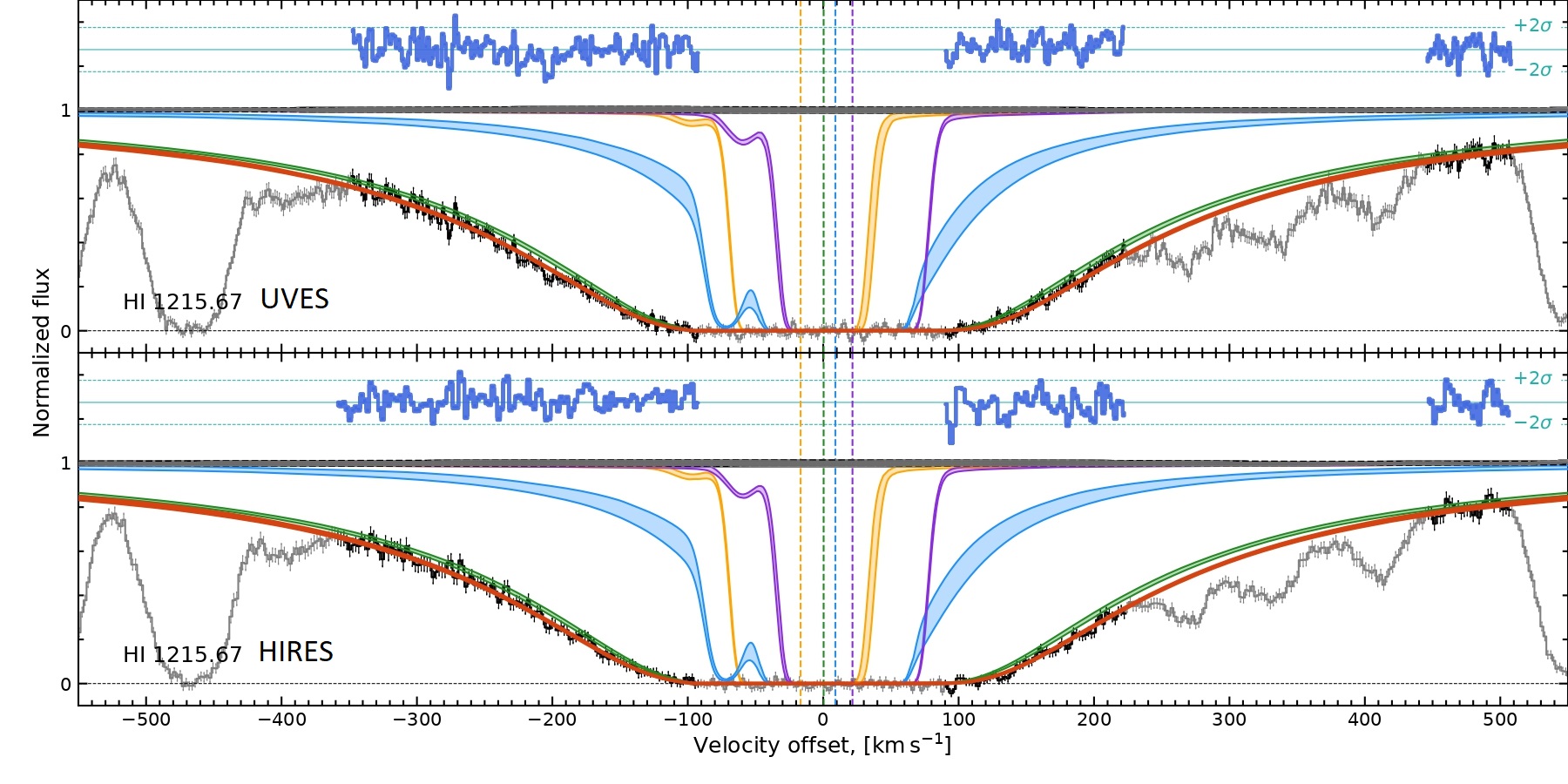}
    \caption{Fit to the \HI\ Ly-$\alpha$ line at $z_{\rm abs}=3.42$ towards \q. The black error bars show the pixels of the observed spectra used in the fit. Other pixels are displayed in grey.
    The total model profile is shown by red lines, while the other coloured lines indicate the contribution of individual components.
    The filled regions between lines correspond to 0.68 credible intervals drawn from the posterior parameter distribution estimated during MCMC fitting.
    Fit residuals are displayed above the profiles.
    }
    \label{fig:Lya}
\end{figure*}

\begin{figure}
	\includegraphics[width=\columnwidth]{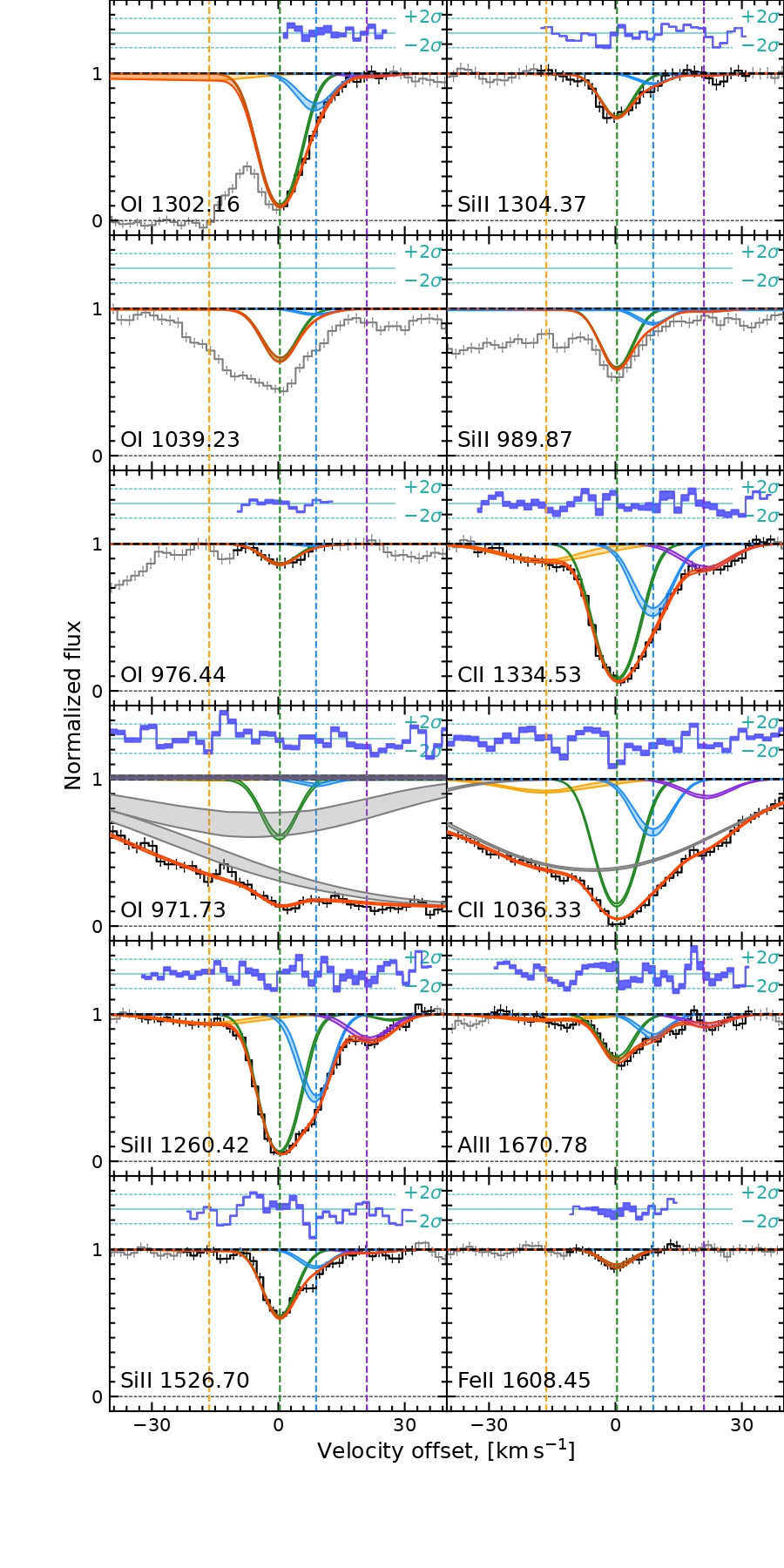}
    \caption{Fits to the metal absorption lines at $z=3.42$ towards \q \ in the UVES spectrum. The graphical conventions of Fig.~\ref{fig:Lya} are used. Transition lines are labelled in each panel. 
    }
    \label{fig:metals1}
\end{figure}

\begin{figure}
	\includegraphics[width=\columnwidth]{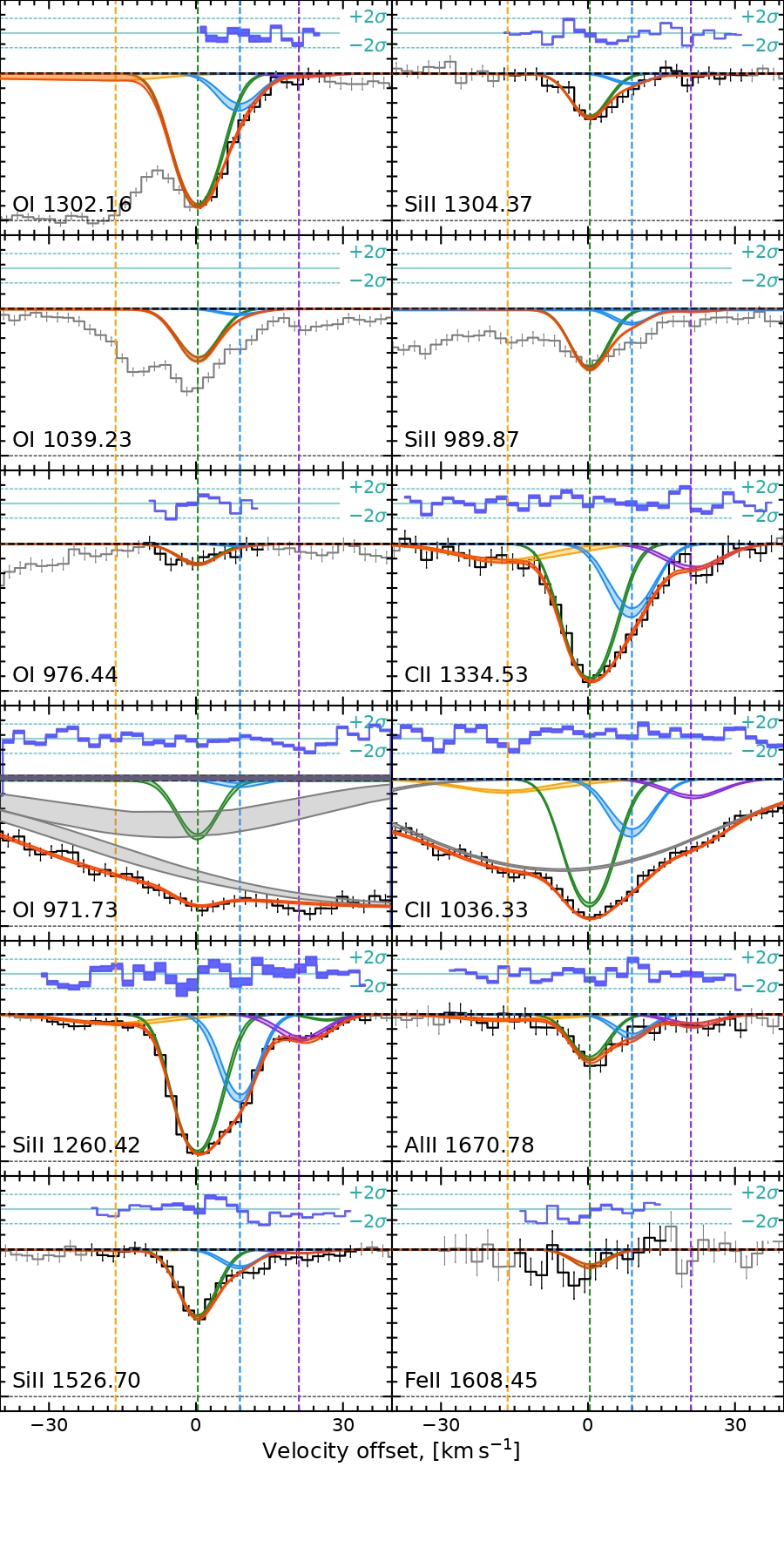}
    \caption{Fits to the metal absorption lines at $z=3.42$ towards \q \ in the HIRES spectum. The graphical conventions of Fig.~\ref{fig:Lya} are used. Transition lines are labelled in each panel. 
    }
    \label{fig:metals2}
\end{figure}

\begin{figure*}
	\includegraphics[width=2\columnwidth]{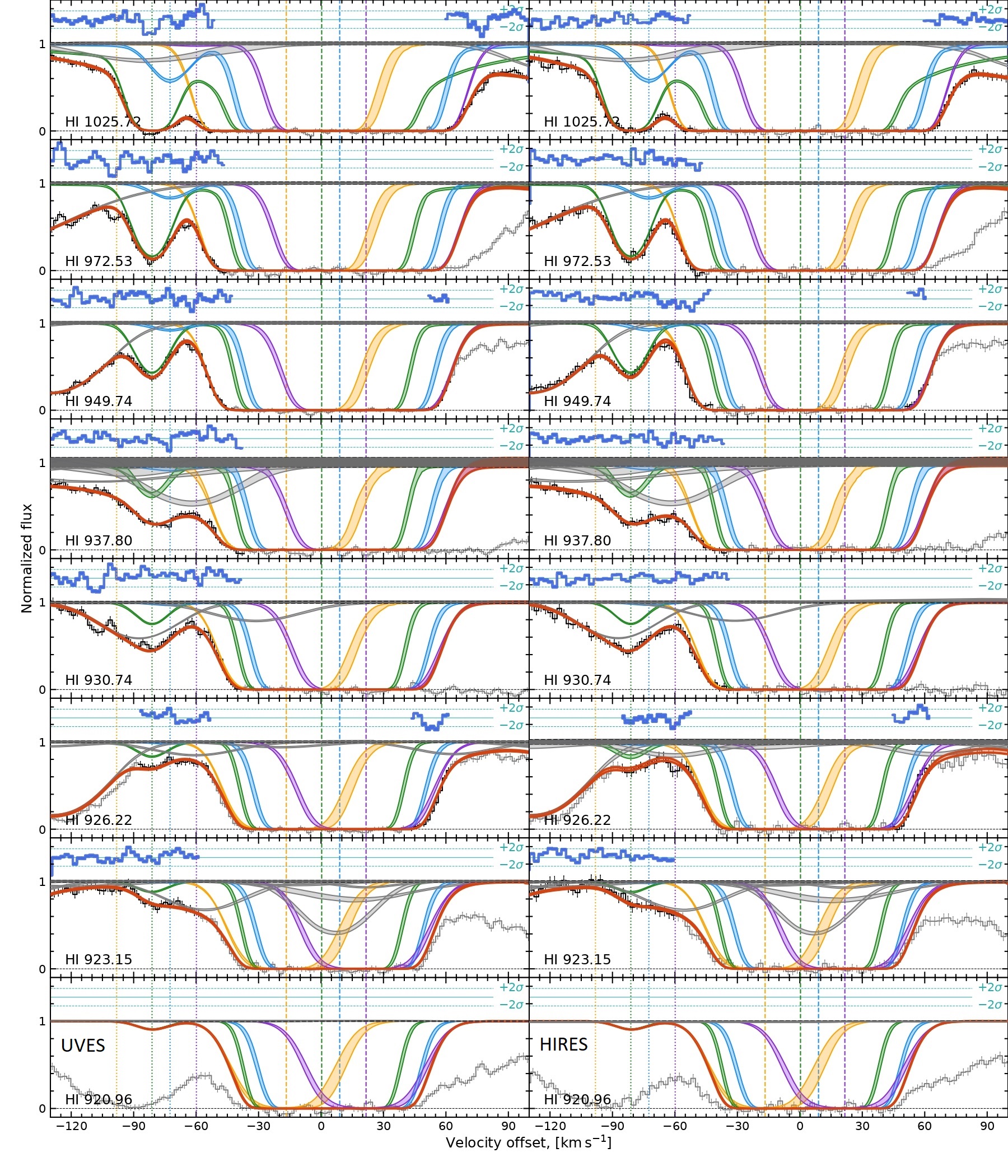}
    \caption{Fits to the \HI\ and \DI\ Lyman-series lines at $z_{\rm abs}=3.42$ towards \q. The left panels show the UVES spectrum while the right panels show the HIRES spectrum. The graphical elements are the same as in Fig.~\ref{fig:Lya}. Transition lines are labelled in each panel.
    }
    \label{fig:D}
\end{figure*}

We used a four-component model to fit \HI, \DI, and low-ionisation metal lines. While the absorption profile is dominated by two main components, with a velocity separation smaller than 10~km\,s$^{-1}$, we added two weak components within a $\approx 40$\,km\,s$^{-1}$ span, barely seen in the strong \CII\,1334 and \SiII\,1260 transition lines. The line profiles of \HI\ Ly$\alpha$, low-ionisation metal species, and the higher-order Lyman series of \HI\ and \DI\ are presented in Figs.~\ref{fig:Lya}, \ref{fig:metals1}, \ref{fig:metals2}, and \ref{fig:D}. The values of the fitting parameters are given in Table~\ref{tab:fit_results}. In the figures, we demonstrate the portions of UVES and HIRES spectra near the sub-DLA absorption system with the best-fit model and the confidence intervals for all components inside the sub-DLA system. The first thing that could catch the eye is a high signal-to-noise ratio which we consider the main source of the small statistical error of our D/H determination. Another point is that despite individual components having noticeable uncertainties shown by the width of colourful lines, the total fit shown in red is very well constrained. To make the analysis more robust we examined the residuals of the fit which we also show in these Figures. In general, the residuals comfortably lie in the range inside 2$\sigma$ with rare outliers in good agreement with the statistical predictions. There are, however, some regions where one could see a quite significant structure and this is also discussed in Sect.~\ref{sect:syst}. We also note that the estimated resolving powers of UVES and HIRES spectra are $R_{\text{UVES}}=50800\pm 1400$, $R_{\text{HIRES}}=49200\pm1500$. These values are in excellent agreement with the nominal values for these spectrographs, and the posterior distribution function indicates no correlation between the D/H value and resolution.

The total \HI\ column density is $\log N=19.304\pm0.004$, indicating that this system belongs to the class of sub-DLAs, where one may expect substantial ionisation corrections. However, thanks to almost the same ionisation potentials of Hydrogen and Deuterium these ionisation corrections are not suitable for the problem of the D/H determination. It has been also shown that the D/H ionisation correction is typically an order of magnitude smaller than current uncertainties of D/H determinations even for sub-DLA systems \citep{Cooke_IC}.

Since Oxygen has an ionisation potential close to the ionisation potential of Hydrogen too, our constraint on the \OI\ column density could be used to derive the metallicity of the system: $\rm [O/H]=-1.71\pm 0.02$. Other metal column densities, such as \CII, \SiII,  \AlII and \FeII\ have values which are consistent with the Oxygen metallicity with small depletion factors (dust content), which is expected at low metallicities. 

From the simultaneous fit of the UVES and HIRES spectra, we find $\log (N($\DI$)/N($\HI$))=-4.623 \pm 0.008$. The quoted error is the formal statistical uncertainty from the Bayesian analysis.

\renewcommand*{\arraystretch}{1.4}

\begin{table*}
\begin{tabular}{ccccccc}
\hline
\hline
Comp. & \#1 & \#2 & \#3 & \#4 & $\log N_{\rm tot}$ & $\rm [X/H]_{total}$ \\
\hline
$\Delta$v, km/s & $-17.4^{+1.1}_{-0.8}$ & $0$ & $8.7^{+0.2}_{-0.2}$ & $21.3^{+0.4}_{-0.4}$ & & \\$b_{\rm turb}$, km/s & $11.7^{+1.9}_{-1.2}$ & $2.8^{+0.1}_{-0.1}$ & $1.0^{+0.5}_{-0.5}$ & $4.2^{+0.8}_{-1.1}$ & & \\ 
$T$, K & $12600^{+1900}_{-2300}$ & $10100^{+200}_{-200}$ & $15100^{+900}_{-700}$ & $22200^{+1200}_{-900}$ & & \\ 
$\log N$(\CII) & $12.83^{+0.05}_{-0.04}$ & $13.86^{+0.02}_{-0.02}$ & $13.18^{+0.03}_{-0.03}$ & $12.75^{+0.04}_{-0.03}$ & $14.003^{+0.01}_{-0.011}$ & $-1.728^{+0.01}_{-0.013}$\\ 
$\log N$(\SiII) & $11.59^{+0.06}_{-0.05}$ & $13.05^{+0.01}_{-0.01}$ & $12.33^{+0.03}_{-0.03}$ & $11.74^{+0.03}_{-0.03}$ & $13.157^{+0.009}_{-0.008}$ & $-1.656^{+0.009}_{-0.01}$ \\ 
$\log N$(\AlII) & $11.1^{+0.09}_{-0.11}$ & $11.66^{+0.03}_{-0.02}$ & $11.29^{+0.04}_{-0.05}$ & $11.06^{+0.08}_{-0.08}$ & $11.949^{+0.017}_{-0.017}$ & $-1.806^{+0.019}_{-0.016}$ \\ 
$\log N$(\FeII) &  & $12.65^{+0.05}_{-0.04}$ & $10.65^{+0.5}_{-0.59}$ & & $12.664^{+0.044}_{-0.047}$ & $-2.140^{+0.043}_{-0.048}$ \\ 
$\log N$(\OI) & $11.77^{+0.49}_{-0.57}$ & $14.25^{+0.02}_{-0.02}$ & $13.2^{+0.05}_{-0.06}$ & $12.13^{+0.13}_{-0.14}$ & $14.29^{+0.02}_{-0.02}$ & $-1.705^{+0.021}_{-0.022}$ \\ 
$\log N$(\HI) & $16.68^{+0.06}_{-0.05}$ & $19.25^{+0.01}_{-0.01}$ & $18.39^{+0.05}_{-0.09}$ & $16.8^{+0.09}_{-0.07}$ & $19.304^{+0.004}_{-0.004}$ & \\ 
\hline 
$\log (N($\DI$)/N($\HI$))$ & $-4.623^{+0.008}_{-0.008}$ &  &  &  &  & \\ 
\hline 
 $\rm[O/H]_{individual}$ & $-1.61^{+0.49}_{-0.58}$ & $-1.69^{+0.03}_{-0.03}$ & $-1.86^{+0.06}_{-0.07}$ & $-1.39^{+0.16}_{-0.15}$ &  & \\
\end{tabular}
\caption{Overall fit results for the absorption-line system at $z=3.4210863(15)$ towards \q. These results correspond to the base model
(see sec.~\ref{sect:method}). We used values from \protect\cite{Asplund_2009} 
for solar abundances: $12 + \log(\text{C/H}) = 8.43$,\ $12 + \log(\text{Si/H}) = 7.51$,\ $12 + \log(\text{Al/H}) = 6.45$,\ $12 + \log(\text{Fe/H}) = 7.50$,\ $12 + \log(\text{O/H}) = 8.69$
}
\label{tab:fit_results}
\end{table*}

\begin{figure*}
	\includegraphics[width=2\columnwidth]{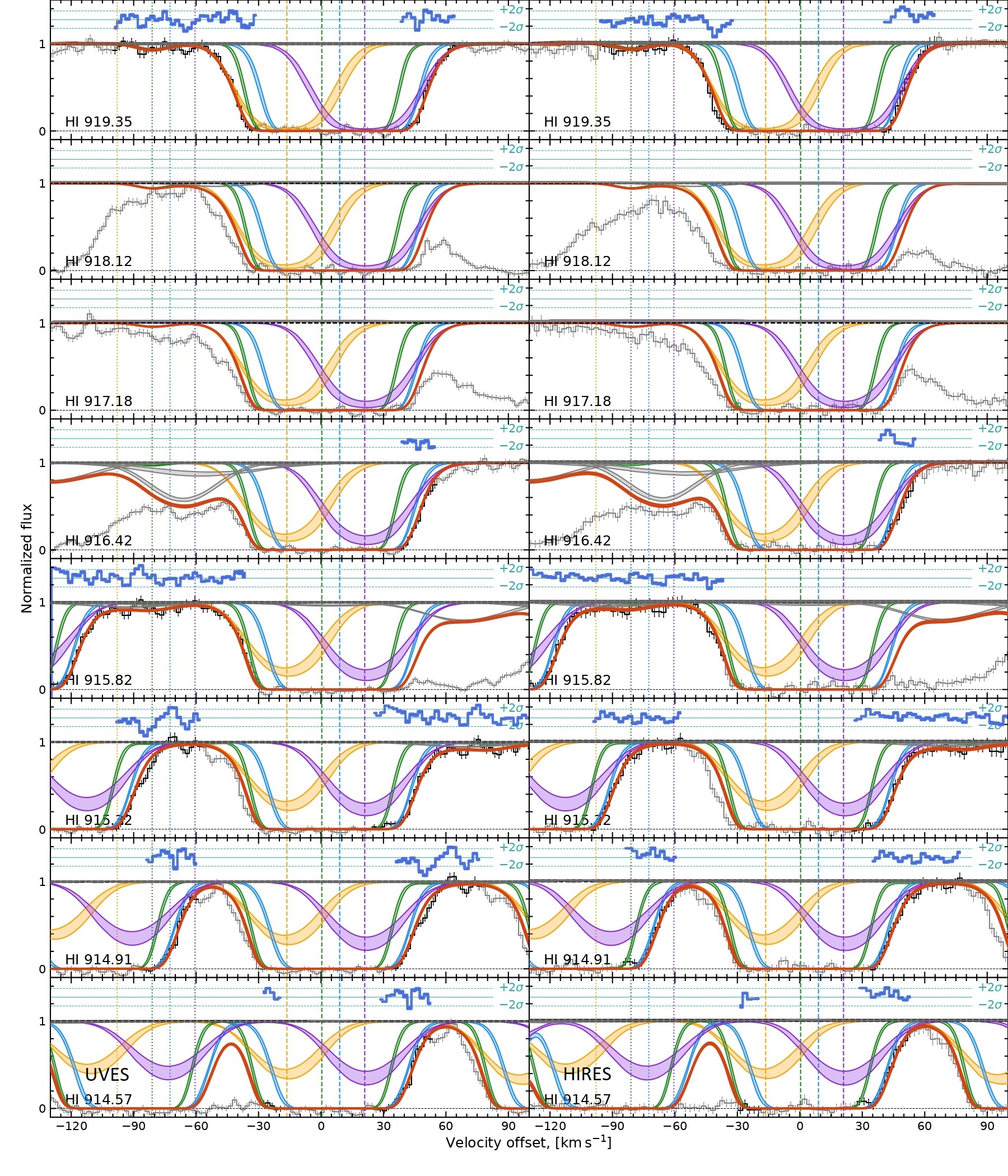}
    \caption{Fits to the \HI\ and \DI\ Lyman-series lines at $z_{\rm abs}=3.42$ towards \q\ (continuation of Fig.~\ref{fig:D}). } 
    \label{fig:D2}
\end{figure*}

\subsection{Systematic uncertainty}
\label{sect:syst}

\begin{figure*}
    \includegraphics[width=2\columnwidth]{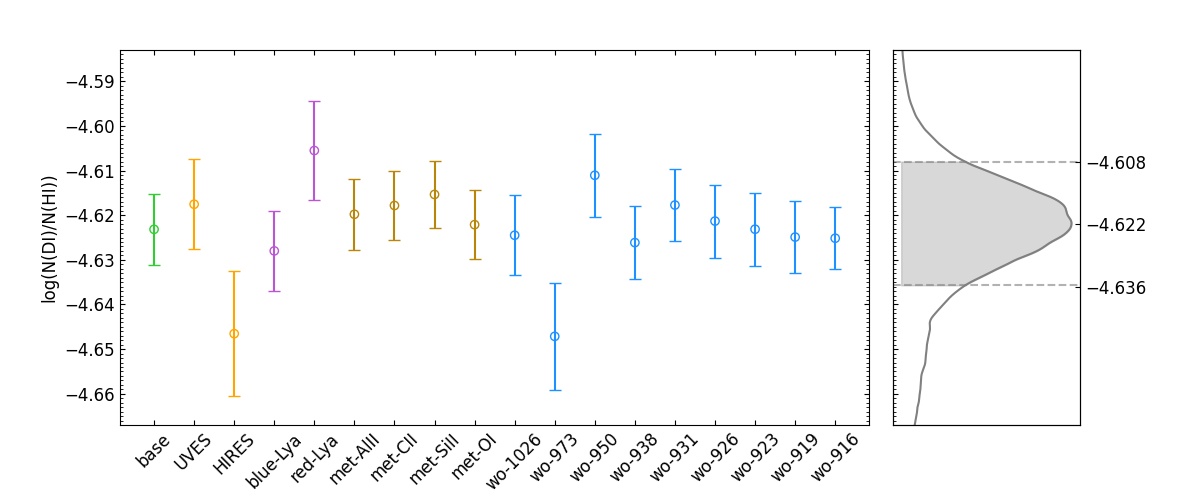}
    \caption{\DI/\HI\ estimates using different model assumptions or spectra (left panel) and joint estimate (right panel).
    Different keywords along the x-axis denote a modification of the base model (labelled as "base" and described in Sect.~\ref{sect:fit}) as follows: "UVES" and "HIRES": only UVES or HIRES is fitted; "blue-Ly$\alpha$" and "red-Ly$\alpha$": only the bluer or the redder wing of the \HI\ Ly$\alpha$ line is fitted; "met-AlII", "met-CII", "met-SiII" and "met-OI": $N($\HI$)$ is tied to either $N($\AlII$)$, $N($\CII$)$, $N($\SiII$)$, or $N($\OI$)$ along the components; "wo-$\lambda$": the region near the \DI\ "$\lambda$" (in \AA) line is excluded from the fit.
    }
    \label{fig:systematic_error}
\end{figure*}

The results derived in the previous section only reported a statistical uncertainty since they were obtained using the specific model assumptions described in Sect.~\ref{sect:method}. To understand the impact of the chosen model assumptions and estimate systematic uncertainties, we subsequently changed each model assumption and performed independent fits. The results are summarized in Fig.~\ref{fig:systematic_error} where we plot the D/H ratios derived through each modification and compare them with our base model.

Using blue symbols in Fig.~\ref{fig:systematic_error}, we show models where we excluded a region near each \HI\ line. These determinations are quite consistent and only the model without a region near \HI\,973\AA\ line is  a little bit outstanding. 
Using olive symbols in Fig.~\ref{fig:systematic_error}, we present the model, where we changed the assumption on the metallicity through the components of the sub-DLA system. We used different elements as a metallicity tracer, as well as fitted the spectrum with a model without this assumption, and we found no significant changes in the D/H ratio in comparison with the base model. By purple colour in Fig.~\ref{fig:systematic_error}, we plot the models where we use only a wing of \HI\,Ly$\alpha$ line. These models indicate typical systematic uncertainty concerned with the measurement of N(\HI).

\begin{figure*}
    \includegraphics[width=2\columnwidth]{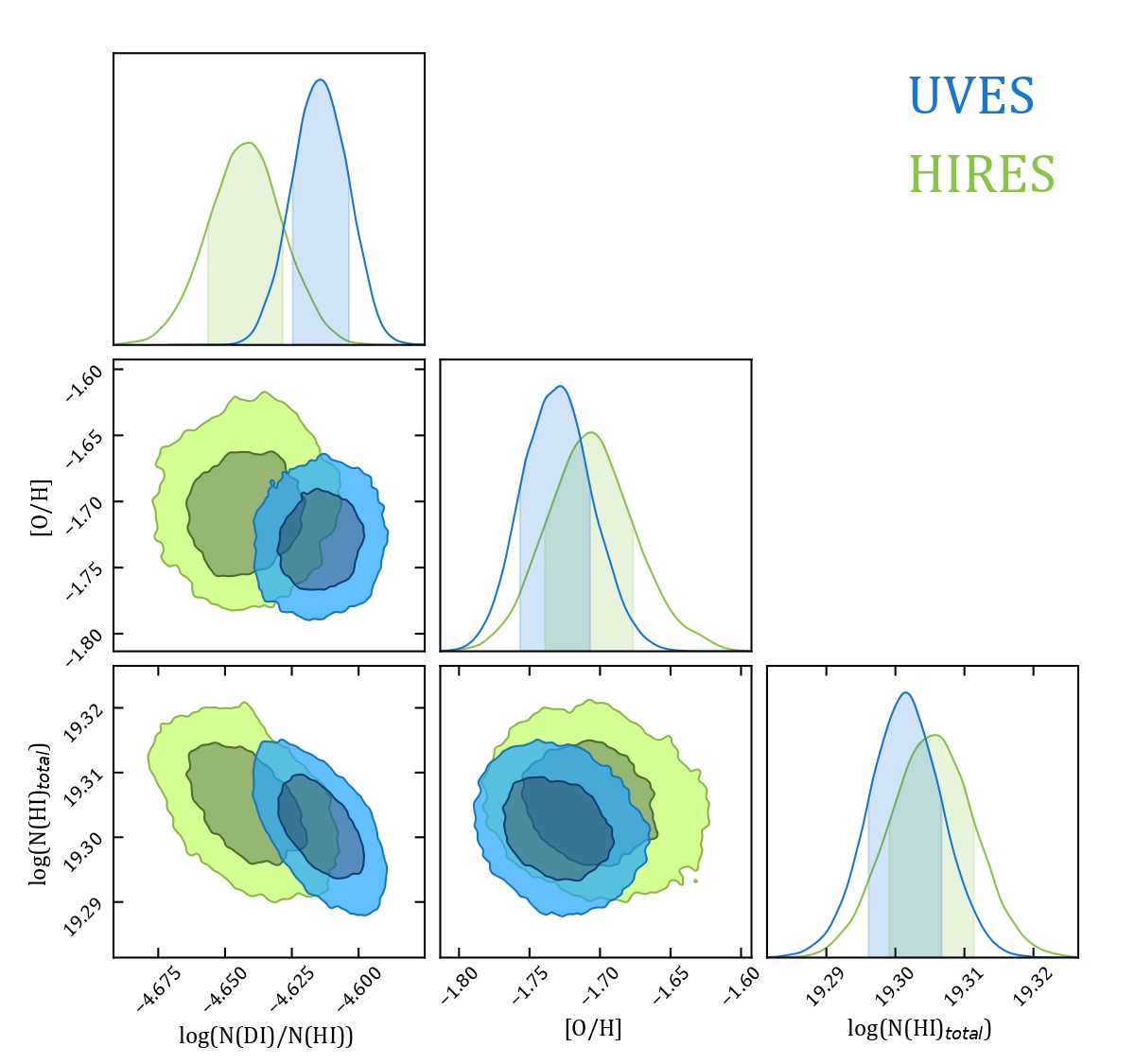}
    \caption{Corner plots showing the posterior distributions of the main parameters in the fitting chains applied to the UVES and HIRES spectra.
 }
    \label{fig:corner}
\end{figure*}

We also fitted VLT/UVES and Keck/HIRES spectra independently and the results are shown in Fig.~\ref{fig:systematic_error} by orange error bars. Posterior distributions of the main parameters are presented in Fig.~\ref{fig:corner}. These $N($\DI$)/N($\HI$)$ determinations differs by $\approx\!\!1.7 \sigma $, which is quite unexpected. As could be seen in Fig.~\ref{fig:corner}, determinations of the total $\log N($\HI$)$ and [O/H] (as well as other nuisance parameters not shown in the figure) are more consistent between two fits, so the main source of the discrepancy is likely the difference in $\log N($\DI$)$.

\begin{figure}
    \includegraphics[width=1\columnwidth]{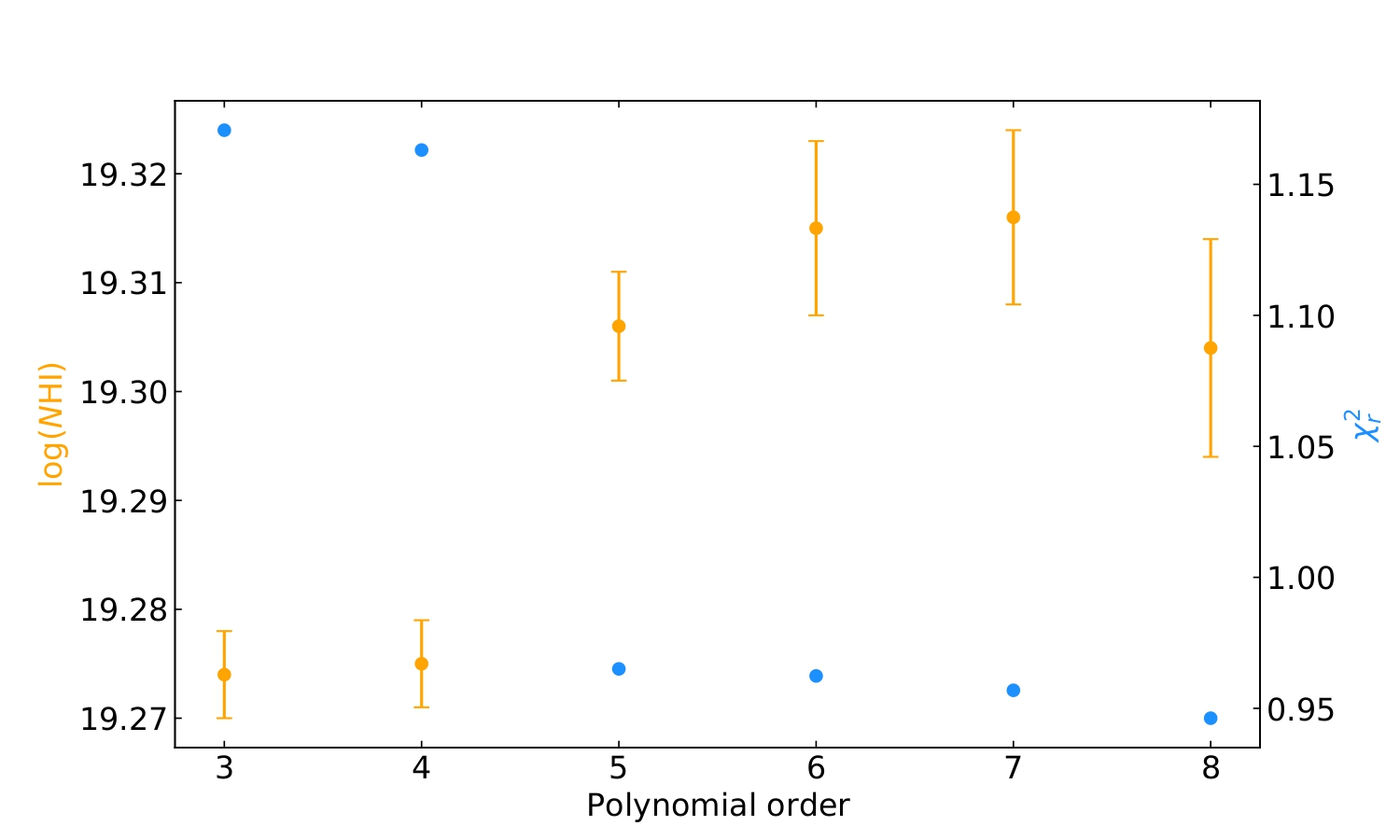}
    \caption{Dependence of the total measured $\log N$(\HI) on the order of the polynomial used for continuum fitting. The orange dots and error bars show the estimated $\log N$(\HI) using a one-component model for Ly$\alpha$ absorption (left ordinate axis). The blue dots show the reduced $\chi^2$ of the fits (right ordinate axis).}
    \label{fig:Polynomial_order}
\end{figure}

We additionally test how the order of the polynomial used for continuum fitting influences the estimated $\log N$(\HI). To do this, we fit only the region near Ly$\alpha$ with different polynomial order for the continuum and only a single component for the Ly$\alpha$ absorption feature. The results are presented in Fig.~\ref{fig:Polynomial_order}. Orders of $3$ and $4$ are too low to fit the spectrum in the region while the higher orders fit the data quite well and do not show the significant scatter in the estimated $\log N$(\HI). Therefore, we choose five as a compromise between the unacceptably simple model and the overfitting. 

We then combined all posterior distribution functions (estimated by MCMC chains) of D/H ratios for these models into one and defined the uncertainty of the $N($\DI$)/N($\HI$)$ determination of $0.014$ dex directly from the final posterior distribution (see Fig.~\ref{fig:systematic_error}). It consists of both statistical and systematic uncertainty.

Another potential source of systematic uncertainty is the fact that there are some regions in the UVES spectrum where noise does not seem to be Gaussian. One could recognise additional absorption features in such a portion of the spectrum like near the \DI\,$\lambda$973 line. For all these cases, comparison with the same regions in the HIRES spectrum mainly shows that there is no demonstration of any additional absorption in the HIRES spectrum and thus we suppose that such a noise is an artefact of the procedure of the UVES spectrum reduction and hence we did not fit all these tiny features as real absorption systems. However, given a noticeable deviation of the D/H value obtained with the model where we exclude \DI\,973$\rm \AA$ line, from the the majority of other models (see Fig.~\ref{fig:systematic_error}), we also fitted all \DI\ lines but $973$\ \AA \  for UVES and HIRES spectra independently to test how fluctuations near \HI\ $973\AA$ line in the UVES spectrum may affect to the estimated \DI/\HI\ value. Using both spectra we estimated $\log$(\DI/\HI)$=-4.647\pm0.012$, while for UVES and HIRES spectra only we got $\log$(\DI/\HI)$=-4.647\pm0.017$ and $\log$(\DI/\HI)$=-4.665\pm0.022$, respectively. It allows us to suppose that the fluctuations near \HI\ $973\AA$ line is not related to the relatively low \DI/\HI\ value for fit without \DI\,$973\AA$, since the \DI/\HI\ value from the fit of the HIRES spectrum alone is even lower than from the fit of the UVES one. 

As a concluding remark on Fig.~\ref{fig:systematic_error}, we would like to highlight two key points. Firstly, all the presented D/H values are demonstrated as alternatives to the base model, and we maintain neutrality among these alternatives; there is no preference for any specific option. Upon analysing the results in the figure, it becomes apparent that a majority of the D/H values tend towards the "UVES" value or suggest that the \DI\ 973 $\AA$ line plays a decisive role, contributing to a higher D/H value. It is crucial to note that, in all cases except "UVES" and "HIRES", we fitted \textit{both} spectra. Our base model involves fitting \textit{all} \DI\ lines in \textit{both} spectra, without favouring any particular metal element and considering \textit{both} wings of the \HI\ Ly$\alpha$ line. Secondly, any further analysis is against the "blind" methodology, as explained in Sect.~\ref{sec:confirmation_bias}. There is no evident reason to favour one spectrum over another, deem one \DI\ line superior to others, or assert the superiority of a specific wing of the \HI\ Ly$\alpha$ line. Therefore, we amalgamate all the results to present the final posterior distribution of the D/H value and its associated error.

We also checked another fit model where an additional \HI\ component was added with the velocity offset of $-81.5$~km\,s$^{-1}$ relative to the major component so \HI\ lines of this additional component were overlapped with \DI\ lines. Obtained result is $N(\mbox{\DI})/N(\mbox{\HI}) = -4.685 \pm 0.014$ (only statistical uncertainty). In principle, there is always such opportunity for any considered D/H measurement, therefore, while we did not include this uncertainty in our final results, we need to bear in mind that the result may be biased. Thus the value of $\Delta = -0.06$ could be perceived as a bias that might take place in the case of 
\DI\ lines that are blended with an unmodelled \HI\ component. 

Lastly, we discussed the uncertainty, related to the profiles of  interloping absorption systems, which may partially blend the total fit near sub-DLA \DI\ lines. In Appendix~\ref{sect:app_A}, we show a set of figures which represent the fit to spectral regions around \DI\ lines in the UVES spectrum. In each figure, we present all the lines for each additional \HI\ component that are available for fitting and were used during the fit. Looking at, e.g., Figs.~\ref{fig:blend_949} and \ref{fig:blends_931}, one can notice that all features near the \DI\,$\lambda$931 line are well constrained by clear absorption by other lines of those absorption systems. The same is relevant for the \DI\,$\lambda$950 line. With such a reliable constraint on the blending absorption features, we suppose that there is little bias of the usage of the partially blended \DI\ lines, as we did in this work. 

\section{Discussion}

The system presented here allowed us to robustly determine the D/H ratio thanks to clear damping wings in Ly-$\alpha$, especially in the regions where the flux immediately recovers from zero (DLA core), which is necessary to accurately constrain the total \HI\ column density in gas-rich (sub-DLA and DLA) systems. Several \DI\ lines are also detected, providing strong constraints on the \DI\ column density. Finally, this particular sub-DLA system exhibits a quite simple velocity structure.
With such a precise measurement, one can discuss the possible astration of deuterium, which is easily destroyed in stars.
Chemical evolution models predict little deviation of D/H from the primordial abundance in predominantly unprocessed gas, i.e., at low metallicity (e.g., \citealt{Dvorkin_2016}). \cite{Weinberg_2016} predicted an astration smaller than 2\% at [O/H$]\approx -1.0$ while similar results were presented by \cite{Van_de_voort_2018}. The latter authors found a mean deviation of D/H from the primordial value in simulations that is smaller than 2\% up to [O/H$]\approx -0.6$.
The system studied here offers a new opportunity to observationally test this at a slightly higher metallicity ([O/H$]\approx -1.7$) than previously possible, confirming the lack of detectable dependence of D/H on metallicity, as also noticed by \cite{Cooke_2016}. In short, we showed that astration remains negligible compared to measurement uncertainties for metallicities of up to at least [O/H$]\approx -1.7$. Naturally, larger statistics are needed to observe tiny deviations from the primordial abundance at the level predicted by the models.

Another point worth mentioning is that this new D/H measurement was performed in a sub-DLA system
with outstanding properties compared to previously studied systems.
Indeed, this sub-DLA has the highest redshift amongst systems in the \textit{precision sample}, has the lowest \HI\ column density, and has one of the highest metallicities. These peculiarities make this  measurement rather unique and relax the conditions allowing for a robust D/H determination. This is especially important in view of the next generation of ground-based telescopes (e.g., ELT, TMT, and GMT) which will be able to observe fainter targets \citep[see also the appendix in][]{Cooke_2016}.

One may also be interested in how metallicity varies from one component to another in the sub-DLA system since it can be used to explore variations in metallicity and study the dust content \citep[see, e.g.,][]{metal_variations}. In the last row of the table~\ref{tab:fit_results} we show the [O/H] for individual components derived using the base model. Oxygen is the best-suited element for metallicity determination, since its ionisation potential is very close to the hydrogen one (and so the ionisation correction is negligible) and there are several strong transitions with appropriate wavelengths. [O/H] determinations for individual components differ from the overall oxygen metallicity by no more than 0.3 dex. This is slightly smaller than the metallicity variations proposed by \cite{De_Cia_2021} and is consistent with other estimates of the metallicity variations \citep{Esteban_2022}.

\subsection{Cosmological implementation}


\begin{figure*}
	\includegraphics[width=2\columnwidth]{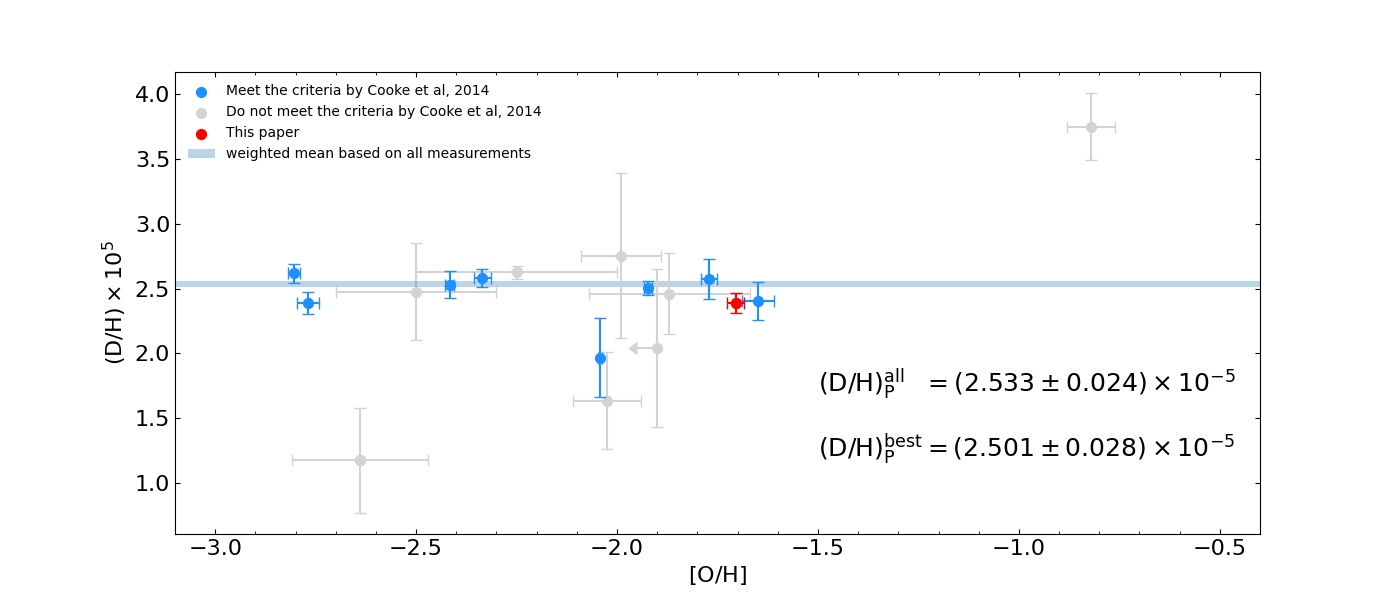}
    \caption{Current D/H measurements. The systems matching the criteria defined by \citet{Cooke_2014} are shown in blue while other determinations are displayed in grey. Our new measurement is shown in red. The weighted mean D/H value based on all measurements is represented by the horizontal line. The weighted mean values based on either all measurements or only the systems matching the above criteria are displayed at the bottom right. 
    }
    \label{fig:met_D_all}
\end{figure*}

The main advantage of the D/H measurements is their direct connection to the primordial deuterium abundance which can be used to infer the cosmological baryon density. Using all available D/H determinations from the literature, which are summarised in table~\ref{tab:previous_determinations}, one can derive the average value (weighted mean):

\begin{equation}
    \rm{(D/H)_{all}} =  (2.533 \pm 0.024)\times 10^{-5}\,.  
    \label{eq:dh_all}
\end{equation}


\begin{table*}
    \centering
    \begin{tabular}{c|c|c|c|c}
        QSO & [O/H] & $\log$(D/H) & reference & meet the selection criteria?\\
    \hline
HS 0105+1619 & -1.771 $\pm$ 0.021 & -4.589 $\pm$ 0.026 & \cite{Cooke_2014} & \checkmark \\
Q0913+072 & -2.416 $\pm$ 0.011 & -4.597 $\pm$ 0.018 & \cite{Cooke_2014} & \checkmark \\
SDSS J1358+6522 & -2.335 $\pm$ 0.022 & -4.588 $\pm$ 0.012 & \cite{Cooke_2014} & \checkmark \\
SDSS J1419+0829 & -1.922 $\pm$ 0.010 & -4.601 $\pm$ 0.009 & \cite{Cooke_2014} & \checkmark \\
SDSS J1558-0031 & -1.650 $\pm$ 0.040 & -4.619 $\pm$ 0.026 & \cite{Cooke_2014} & \checkmark \\
Q1243+307 & -2.769 $\pm$ 0.028 & -4.622 $\pm$ 0.015 & \cite{Cooke_2018} & \checkmark \\
SDSS J1358+0349 & -2.804 $\pm$ 0.015 & -4.582 $\pm$ 0.012 & \cite{Cooke_2016} & \checkmark \\
J1444+2919 & -2.042 $\pm$ 0.005 & -4.706 $\pm$ 0.067 & \cite{Balashev_2016} & \checkmark \\
CTQ 247 & -1.990 $\pm$ 0.100 & -4.560 $\pm$ 0.100 & \cite{Noterdaeme_2012} & $\times$ \\
PKS 1937-1009 (1) & -1.870 $\pm$ 0.200 & -4.610 $\pm$ 0.050 & \cite{Riemer-Sorensen_2015} & $\times$ \\
J1337+3152 & -2.640 $\pm$ 0.170 & -4.930 $\pm$ 0.150 & \cite{Srianand_2010} & $\times$ \\
J1134+5742 & $<$-1.9  & -4.690 $\pm$ 0.130 & \cite{Fumagalli_2011} & $\times$ \\
Q2206-199 & -2.070 $\pm$ 0.050$^{\rm a}$ & -4.786 $\pm$ 0.100 & \cite{Pettini_2001} & $\times$ \\
Q0347-3819 & -0.820 $\pm$ 0.060 & -4.426 $\pm$ 0.028 & \cite{Levshakov_2002} & $\times$ \\
PKS 1937-1009 (2) & -2.250 $\pm$ 0.250 & -4.581 $\pm$ 0.008 & \cite{Riemer-Sorensen_2017} & $\times$ \\
Q1009+2956 & -2.500 $\pm$ 0.200 & -4.606 $\pm$ 0.066 & \cite{Zvarygin_2018} & $\times$ \\
J1332+0052 & -1.725 $\pm$ 0.019 & -4.622 $\pm$ 0.014 & this paper & \checkmark  
        \end{tabular}
        \flushleft $^{\rm a}$ presented in \cite{Pettini_2008}\\
    \caption{Current sample of D/H determinations. In the last column we show either a measurement meets or not the selection criteria of the \textit{precision sample} proposed by \protect\cite{Cooke_2014}.}
    \label{tab:previous_determinations}
\end{table*}

This is our conservative estimate in the sense that it incorporates all previous determinations.
We also show the comparison of the new D/H measurement with the previous ones in Fig.~\ref{fig:met_D_all}. In this figure, we show D/H versus [O/H] for all known determinations and we used blue colour for measurements which meet the selection criteria proposed by \citet{Cooke_2014}, red colour for our new determination and grey colour for all other determinations. The new measurement is consistent with other precision measurements and has a similar uncertainty. Fig.~\ref{fig:met_D_all} also shows that there is no obvious correlation between metallicity and D/H ratio.

Up to now, there is no robust verification that the spread of the D/H determinations is not physical. However, more high-quality observations can clarify estimates as shown by \cite{Cooke_2014}. Moreover, these authors offered a set of criteria to select the most suitable cases for D/H determinations where some systematic effects are negligible. Based on the nine systems which agree with these criteria (see table~\ref{tab:previous_determinations}), one can obtain a weighted mean of presumably best-suited measurements:

\begin{equation}
    \rm{(D/H)_{best}} =  \left(2.501 \pm 0.028 \right) \times 10^{-5}\,.
    \label{eq:dh_best}
\end{equation}     

Even if measurements that meet the selection criteria of the \textit{precision sample} are presumed to be the most suitable for D/H determination, we suggest that it is important to use all available data and measurements made by different researchers, and using different techniques, until this presumption is demonstrated conclusively.

Both values (eq. \ref{eq:dh_all}, \ref{eq:dh_best}) are
in moderate tension ($\approx\!\!2.2\sigma, \ \approx\!\!1.4\sigma$ respectively) with the theoretically-predicted value of the primordial deuterium abundance computed using the PRIMAT code and based on the cosmological parameters derived from CMB+BAO \citep{PL18} presented by \cite{Pitrou2021}:

\begin{equation}
    (\rm{D/H})_{\rm th} =  \left(2.439 \pm 0.037 \right) \times 10^{-5}
\end{equation}    
assuming $\tau_n = 879,4(6) s $ and  $100h^2
\Omega_b = 2.242 (\pm0.014) $. An important and significant achievement in the $\rm D(p,\gamma)^3He$ rate determination was reached in the last years by the Laboratory for Underground Nuclear Astrophysics (LUNA) staff \citep{Mossa_2020}. This reaction was the main source of uncertainty in predicting the theoretical primordial D/H ratio.
\cite{Pitrou2021} used the new rate of this reaction for the computations and the new rate enhanced the tension. 

This growing discrepancy which was also mentioned by \cite{Cooke_2018} and \cite{Pitrou2021} could be a manifestation of new physics beyond the Standard Cosmological Model.

\subsection{Avoiding confirmation bias}
\label{sec:confirmation_bias}

Given the number of previous works on the primordial D/H ratio measurements as well as BBN+CMB value, one could expect that a new determination should be consistent with others. This could in principle influence the decisions made through the analysis of new D/H systems. 

First of all, it is worth reminding that D/H measurements are only a few so any new measurement is extremely valuable. The detection of a clear \HI\ Ly$\alpha$ line and several optically thin \DI\ lines lead us to make the analysis to the end, regardless of the obtained D/H value. 

While not performing a complete blind analysis (as done by \citealt{Cooke_2014}), where the value of D/H is kept unknown until 
all decisions about the modelling are accepted based on the visual matching of the data by the fit and $\chi^2$ values, we did not favour any D/H value during the fitting process.
In fact, Fig.~\ref{fig:systematic_error} presents only a small scatter in the D/H determinations between our different models. Any prior based on values by, e.g., \citet{Cooke_2018} could be anywhere within a $\sim 0.1$~dex range around $\log ({\rm D/H})=-4.61$, and all our models are comfortably within this range. We would have no reason to subconsciously prefer some value within that range. In other words, a confirmation bias would rather take place for a choice between, e.g., a value of $-4.6$ and $-4.4$, i.e. with a deviation much more substantial than observed here. The best way to avoid subjectivity remains to perform a D/H determination in each system where it is possible to do and confront analysis obtained by different groups and methods.

\section{Conclusions}

To date, there are only 17 determinations of the D/H ratio in metal-poor DLA/sub-DLA systems at high redshift. Nine of them agree with the criteria defined by \citet{Cooke_2014} to qualify as robust measurements. In this way, each new determination is crucial, as well as any reanalysis of known systems based on new high-quality data.

Our main conclusions are:
\begin{enumerate}
    \item Based on high-quality spectra we obtained a new measurement of the  deuterium abundance in the metal-poor sub-DLA system at $z = 3.42$ towards
quasar J\,1332+0052: ${\rm D/H}=(2.388 \pm 0.078)\times 10^{-5}$.
    \item This new measurement is consistent with previous determinations of the same precision level and meets the selection criteria proposed by \cite{Cooke_2014}.
    \item Based on all available D/H measurements we estimated a new primordial deuterium abundance D/H$_{\rm pr}=(2.533 \pm 0.024)\times 10^{-5}$. Based on nine measurements which meet the criteria of a precise one we find D/H$_{\rm pr}=(2.501 \pm 0.028)\times 10^{-5}$. 
    \item We used a new method for accounting the systematic effects based on multiple times fitting, when we do an independent fitting for each model we consider it appropriate for describing the data. The main advantage of this method is that we do not choose which model is better or worse which might be subconsciously biased by our human nature and combine all the results from any possible model instead. Despite it being very time-consuming, it is possibly the single ability to realise how different aspects could affect the results of the analysis. 
    \item A new determination of the D/H$_{\rm pr}$ based on all measurements is still marginally inconsistent ($\approx$2.2$\sigma$) with the theoretical value predicted by \cite{Pitrou2021} based on the cosmological parameters derived from CMB+BAO \citep{PL18}. The reasons for this
    remain unsettled and
    could potentially be an indication of new physics.
    \item There is still no obvious correlation between D/H and [O/H] or N(\HI).
\end{enumerate}

\section*{Acknowledgements}

We thank the anonymous referee for the careful review and useful suggestions.
The authors are grateful to the VLT and Keck observatory staff for acquiring the analysed data
and to Ryan Cooke for his invaluable input on this particular system.
All performance-expensive calculations were performed on the cluster at Ioffe Institute. This work was supported by RSF grant 23-12-00166.

Some of the data presented herein were obtained at Keck Observatory, which is a private 501(c)3 non-profit organization operated as a scientific partnership among the California Institute of Technology, the University of California, and the National Aeronautics and Space Administration. The Observatory was made possible by the generous financial support of the W. M. Keck Foundation. The authors wish to recognize and acknowledge the very significant cultural role and reverence that the summit of Maunakea has always had within the Native Hawaiian community. We are most fortunate to have the opportunity to conduct observations from this mountain. This research has made use of the Keck Observatory Archive (KOA), which is operated by the W. M. Keck Observatory and the NASA Exoplanet Science Institute (NExScI), under contract with the National Aeronautics and Space Administration.

\section*{Data Availability}

The pipeline-reduced spectra underlying this article are readily available from the ESO science portal (\url{http://archive.eso.org/scienceportal/home}) and partially from the Keck observatory archive (\url{https://koa.ipac.caltech.edu}). Re-reduced spectra can be shared upon request to the corresponding author.



\bibliographystyle{mnras}
\bibliography{example} 




\appendix

\section{Interloping systems}
\label{sect:app_A}

\begin{figure*}
    \includegraphics[width=2\columnwidth]{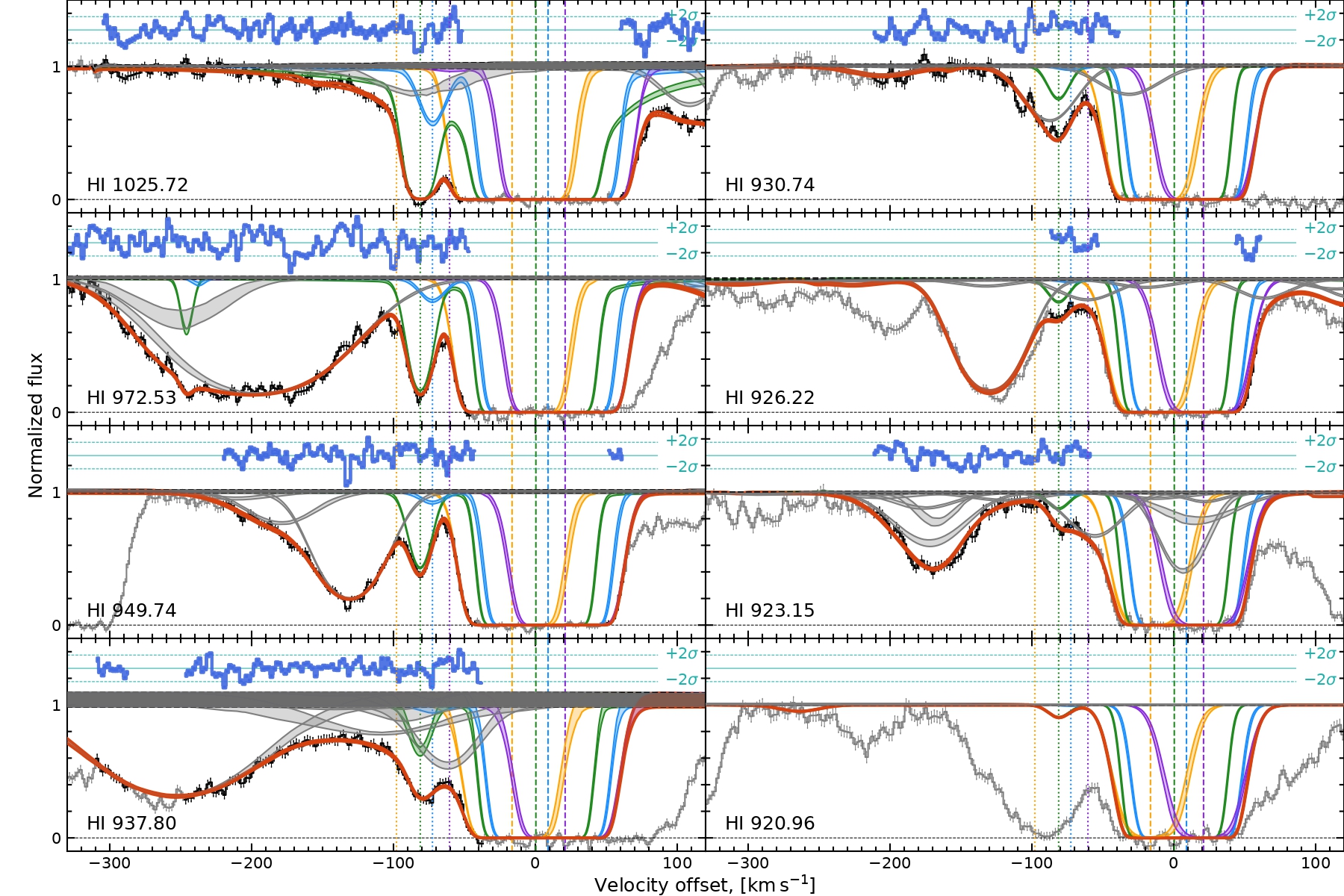}
    \caption{Portion of UVES spectrum with a focus on unrelated absorption features.}
    \label{fig:all_blends}
\end{figure*}

In this section, we present the modelling of unrelated absorption-line systems. 
In figure~\ref{fig:all_blends} we present strongest \DI\ absorption lines of the sub-DLA system and interloping absorption features near them in the UVES spectrum. 
Each following figure shows a portion of the UVES spectrum with the fit 
in more details. In particular, these figures show additional absorption lines which allow one to better constrain parameters of a system. The interloping systems are shown in grey. The other colours are those used in the main figures to either depict the total model profile (red) or individual sub-DLA components (i.e., orange, green, blue, and violet for components \#1, \#2, \#3, and \#4, respectively). \DI\ lines usually appear in green.

\begin{table}
\centering
\caption{Unrelated absorption systems}
\label{}
\begin{tabular}{cc}
\hline
line in sub-DLA system & Blending system Figure \\
\hline
$\lambda1026$ & \ref{fig:blend_1026} \\
$\lambda973$ & \ref{fig:blend_973} \\
$\lambda950$ & \ref{fig:blend_949} \\
$\lambda938$ & \ref{fig:blend_938}, \ref{fig:blend_938_2} \\
$\lambda931$ & \ref{fig:blends_931} \\
$\lambda926$ & \ref{fig:blends_926} \\
$\lambda923$ & \ref{fig:blends_922} \\
\hline
\end{tabular}
\end{table}

\begin{figure*}
    \includegraphics[width=2\columnwidth]{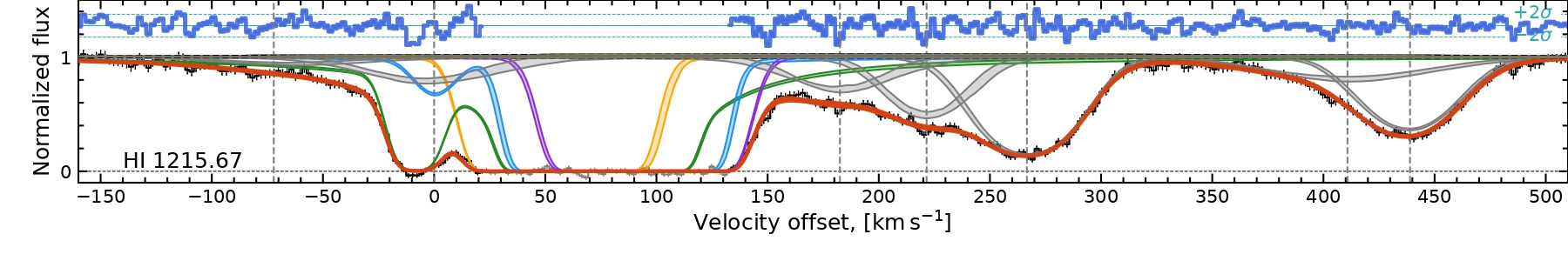}
    \caption{Absorption system interloping with \DI\,$\lambda$1026. 
    The system is located at redshift $z=2.729383$. Other systems which affect total fit near the \HI\ line are also shown.}
    \label{fig:blend_1026}
\end{figure*}

\begin{figure}
    \includegraphics[width=1\columnwidth]{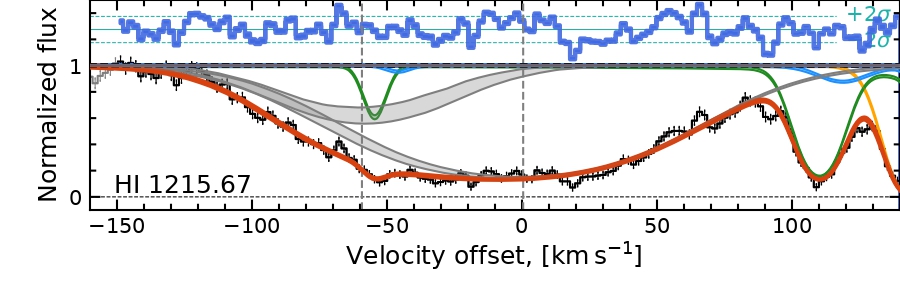}
    \caption{Absorption system interloping with \DI\,$\lambda$973. The system is located at redshift $z=2.534616$. Another system nearby is also shown.}
    \label{fig:blend_973}
\end{figure}

\begin{figure*}
    \includegraphics[width=2\columnwidth]{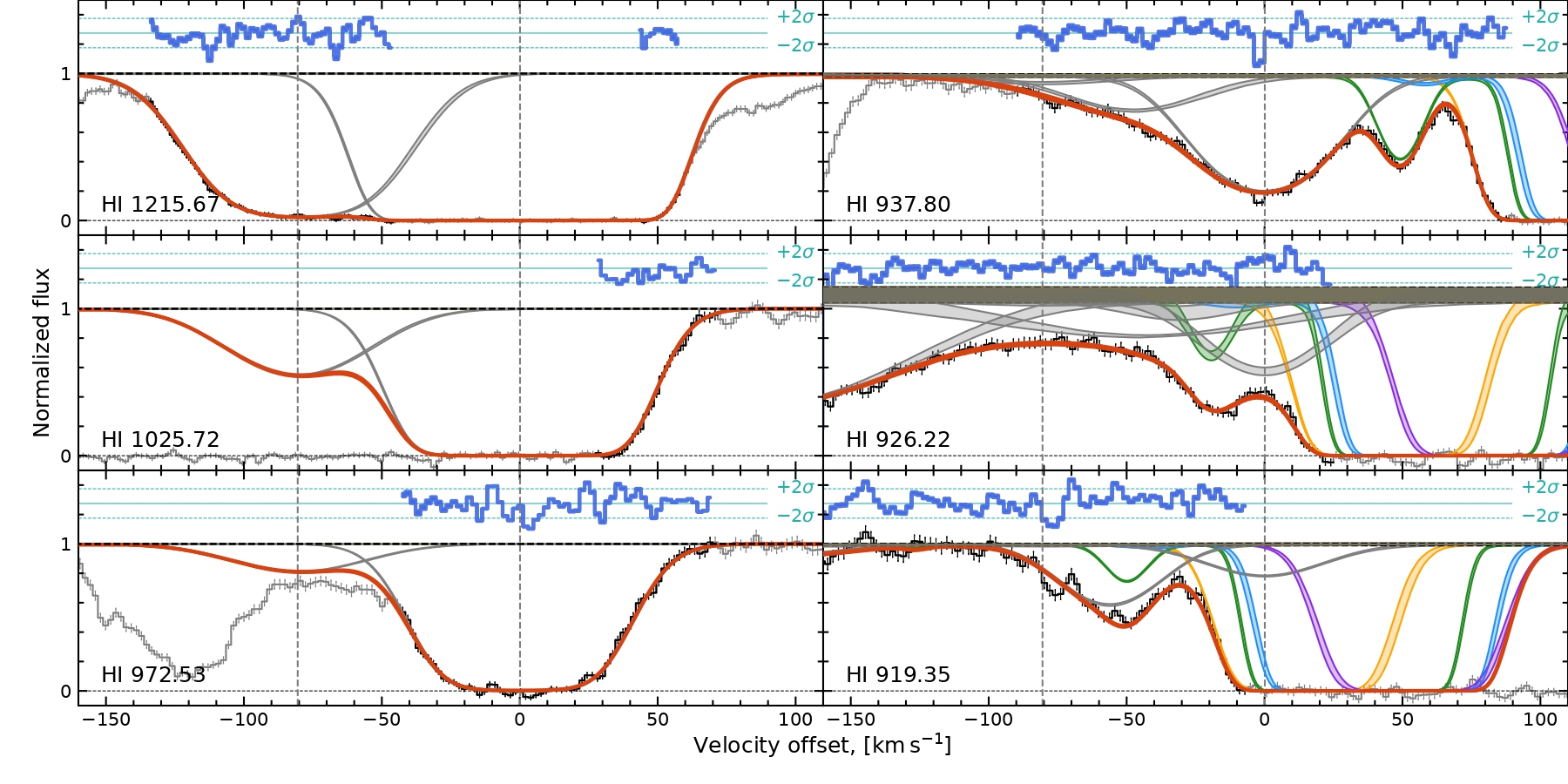}
    \caption{Absorption systems interloping with \DI\,$\lambda$950. The velocity offset is counted relative to the system located at redshift $z=3.475422$.}
    \label{fig:blend_949}
\end{figure*}

\begin{figure}
    \includegraphics[width=1\columnwidth]{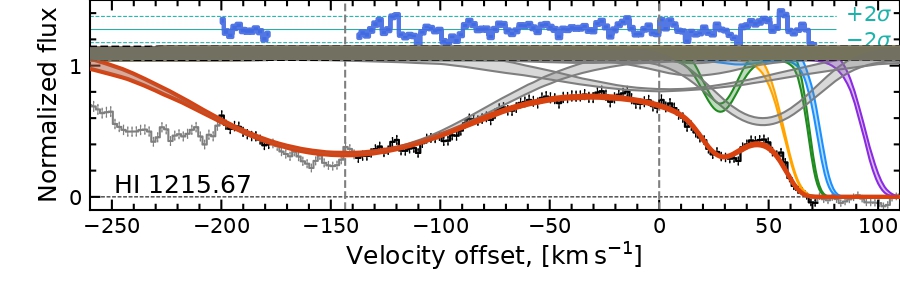}
    \caption{The first absorption system interloping with \DI\,$\lambda$938. The velocity offset is counted relative to the system located at redshift $z=2.409310$.}
    \label{fig:blend_938}
\end{figure}

\begin{figure}
    \includegraphics[width=1\columnwidth]{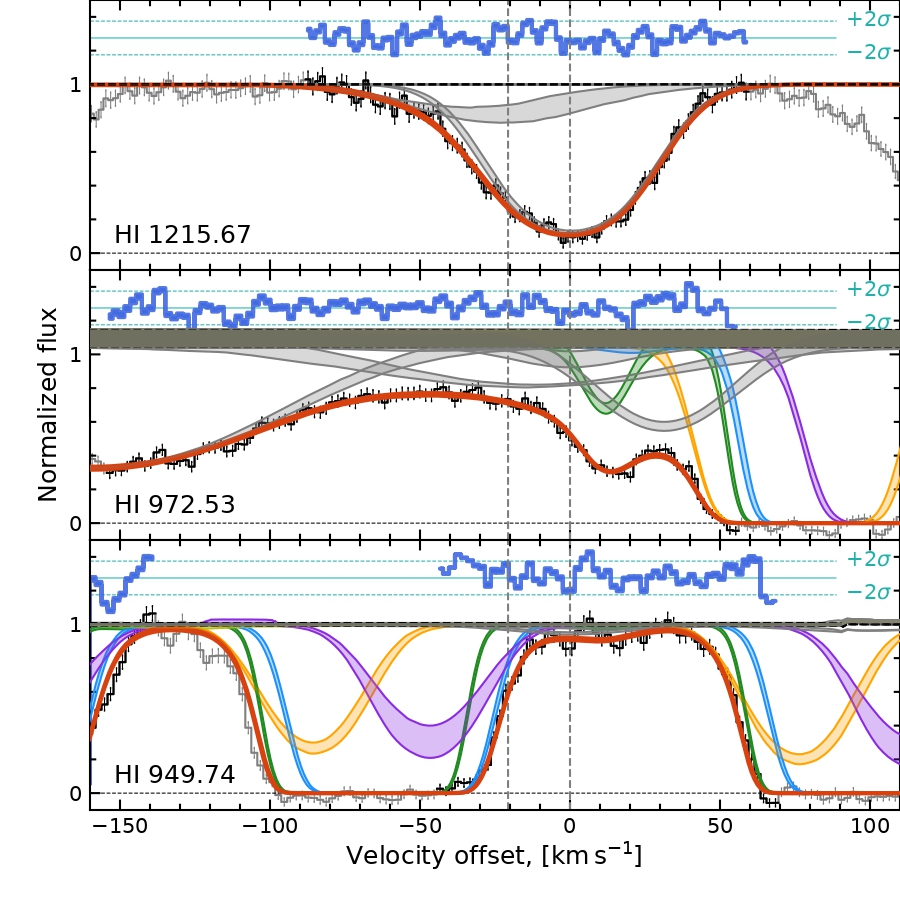}
    \caption{The second absorption system interloping with \DI\,$\lambda$938. The velocity offset is counted relative to the system located at redshift $z=3.261862$.}
    \label{fig:blend_938_2}
\end{figure}

\begin{figure}
    \includegraphics[width=1\columnwidth]{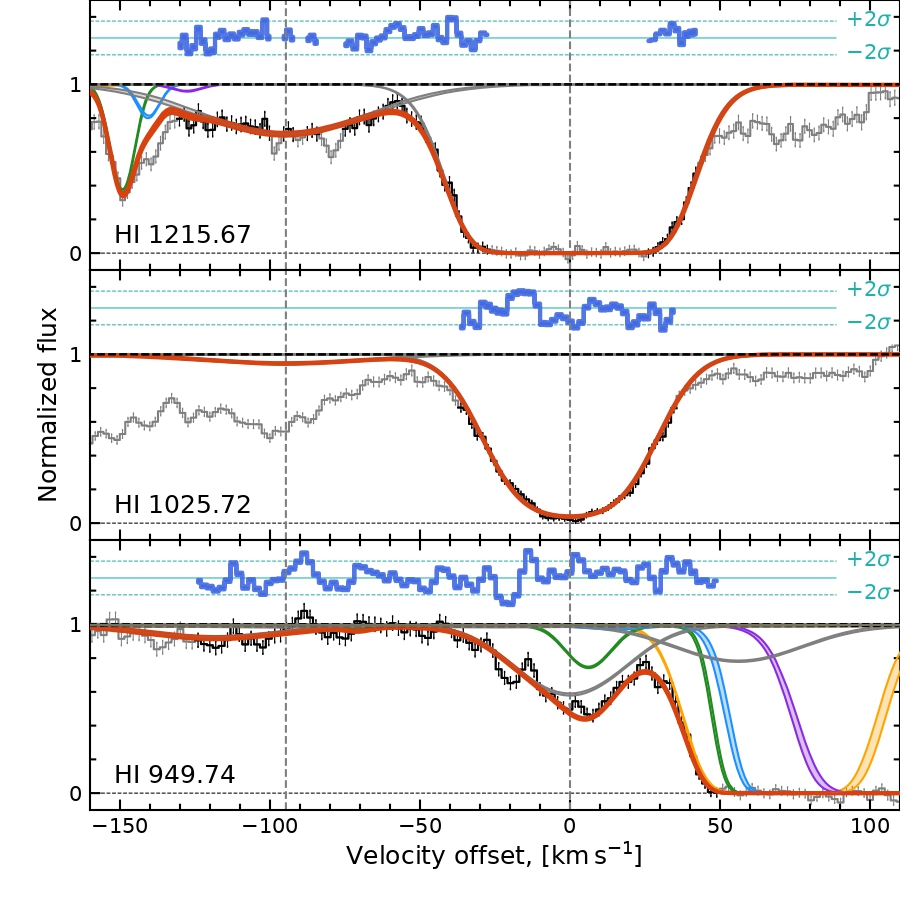}
    \caption{Absorption system interloping with \DI\,$\lambda$931. The velocity offset is counted relative to the system located at redshift $z=3.331398$.}
    \label{fig:blends_931}
\end{figure}

\begin{figure}
    \includegraphics[width=1\columnwidth]{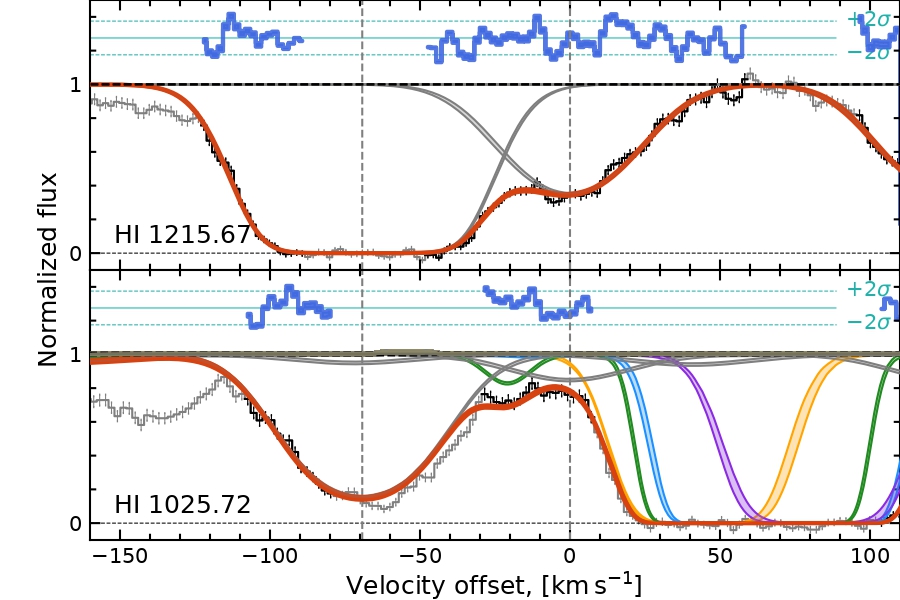}
    \caption{Absorption system interloping with \DI\,$\lambda$926. The velocity offset is counted relative to the system located at redshift $z=2.991426$.}
    \label{fig:blends_926}
\end{figure}

\begin{figure}
    \includegraphics[width=1\columnwidth]{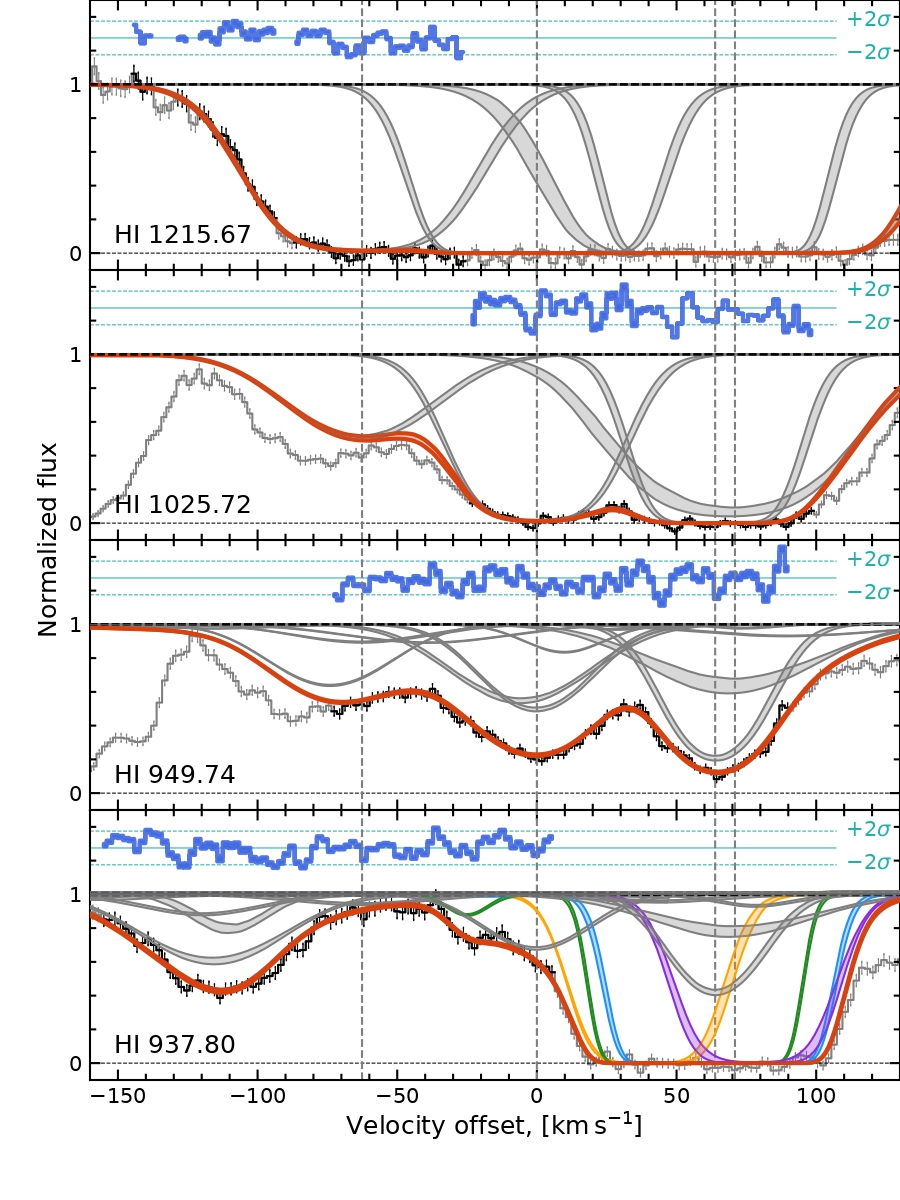}
    \caption{Absorption system interloping with \DI\,$\lambda$923. The velocity offset is counted relative to the system located at redshift $z=3.351185$.}
    \label{fig:blends_922}
\end{figure}


\bsp	
\label{lastpage}
\end{document}